\documentclass{emulateapj}

\begin{document}
\title{Don't Forget the Forest for the Trees:  The stellar-mass halo-mass relation in different environments}
\author{Stephanie Tonnesen$^{1}$ and Renyue Cen$^{2}$}
\affil{$^{1}$Carnegie Observatories, 813 Santa Barbara St, Pasadena, CA, 91101 $^{2}$Department of Astrophysics, Princeton University, Peyton Hall, Princeton, NJ, 08544}
\email{1 stonnes@gmail.com (ST)\\ 2 cen@astro.princeton.edu (RC)} 

\abstract{The connection between dark matter halos and galactic baryons is often not well-constrained nor well-resolved in cosmological hydrodynamical simulations.  Thus, Halo Occupation Distribution (HOD) models that assign galaxies to halos based on halo mass are frequently used to interpret clustering observations, even though it is well-known that the assembly history of dark matter halos is related to their clustering.  In this paper we use high-resolution hydrodynamical cosmological simulations to compare the halo and stellar mass growth of galaxies in a large-scale overdensity to those in a large-scale underdensity (on scales of about 20 Mpc).  The simulation reproduces assembly bias, that halos have earlier formation times in overdense environments than in underdense regions.  We find that the stellar mass to halo mass ratio is larger in overdense regions in central galaxies residing in halos with masses between 10$^{11}$-10$^{12.9}$ M$_{\odot}$.  When we force the local density (within 2 Mpc) at $z$=0 to be the same for galaxies in the large-scale over- and underdensities, we find the same results.  We posit that this difference can be explained by a combination of earlier formation times, more interactions at early times with neighbors, and more filaments feeding galaxies in overdense regions.  This result puts the standard practice of assigning stellar mass to halos based only on their mass, rather than considering their larger environment, into question.}
}
\section{Introduction}

In a $\Lambda$ cold dark matter ($\Lambda$CDM) universe (Komatsu et al. 2011), the mass function and distribution of dark matter halos is well-understood.  Dark matter only simulations show that halos form in filaments that are connected at nodes containing (a) massive halo(s).  This compares well with observations that show galaxies in filaments, sheets, clusters and superclusters (e.g. Springel et al. 2006).  

Connecting the stellar component of galaxies to dark matter halos can be difficult because galaxy growth depends on both internal and external factors.  Gas accretion and mergers can drive galaxy star formation, while feedback from active galactic nuclei (AGN), supernovae and stellar winds can delay or stop growth.  Many of these processes are difficult for cosmological hydrodynamical simulations to resolve, such as the radiative cooling of gas into molecular clouds, star formation, and energy injection into the interstellar medium from supernovae and AGN. 

Despite these complications, halo mass has been closely linked to a galaxy's stellar mass and star formation.  Analytic models of galaxy formation through smooth gas accretion found that there is an upper mass threshold at which accreted gas is shocked to temperatures from which it can no longer cool within a Hubble time and file star formation, determining the mass at which galaxies must be red (Rees \& Ostriker 1977; Silk 1977; Binney 1977; White and Rees 1978).  Recent numerical hydrodynamical cosmological simulations by Kere\v{s} et al. 2005 (see also Ocvirk et al. 2008) have updated this model, showing that cold gas accretion can occur in halos with a range of masses, but the amount of cold gas in a galaxy is determined by its dark matter halo mass, M$_{\mathrm{halo}}$.  Therefore, one would expect a galaxy's M$_{\mathrm{halo}}$ to  be directly linked to the SFR from cold gas accretion.  This model is focused only on the central galaxies in halos, as these are the only galaxies for which we could expect a relationship between gas and dark matter halo mass based on the two-mode theory of gas accretion (Kere\v{s} et al. 2005; Dekel \& Birnboim 2006).  A number of possible interactions can affect the gas, stellar, and dark matter mass of a satellite galaxy, as discussed in Boselli \& Gavazzi (2006). 

Crain et al. (2009) use the GIMIC simulations--hydrodynamical ``re"-simulations of the Millennium simulation at (-2, -1, 0, +1, +2)$\sigma$ of the mean density on the scale of $\sim$20 Mpc--to determine whether halo mass is responsible for the differences in the SFR density (SFRD) in different environments.  They find that the SFR to halo mass ratio is the same in all environments, and conclude that halo mass determines the rate at which galaxies form stars.

Because of the difficulty in directly modeling the baryonic physics that drives gas cooling, star formation, and feedback, and using the theory that halo mass is the fundamental parameter that determines other galaxy properties, more analytic models have been used to connect galaxies to dark matter halos.  In particular, standard Halo Occupation Distribution models (HODs) assume that halo mass is the fundamental parameter determining the stellar mass and internal processes of galaxies, and do not include any dependence on the larger environment in which the halo resides.  These models are frequently used to make mock catalogues to interpret galaxy clustering measurements (e.g. Kauffmann et al. 1997; Jing et al. 1998; Benson et al. 2000; Seljak 2000; Peacock \& Smith 2000; Ma \& Fry 2000; Scoccimarro et al. 2001; Berlind \& Weinberg 2002; Zheng et al. 2005; Tinker et al. 2008).  

Because the halo mass function depends on environment--more massive halos form in higher density regions of the universe, called halo bias (e.g. Kaiser 1984)--these methods also reproduce the empirical finding that higher stellar mass galaxies reside in regions of higher galaxy density.  For example, Abbas \& Sheth (2006) compare galaxy clustering in the SDSS DR4 (M$_{r}$ $<$ -21) to measurements in mock catalogues created using HODs and to measurements from an analytic halo model calculation.  The authors argue that the three samples agree well enough that correlations between galaxy properties can be entirely explained by the variation of the halo mass function in different environments (see also Skibba et al. 2006).  Tinker et al. (2008) compare observed void statistics to those obtained using a standard HOD, and find that the sizes and emptiness of voids are in excellent agreement.

However, more than the halo mass distribution has been shown to depend on the environment.  
Gao et al. (2005) use the Millennium Simulation to find that the $\Lambda$CDM paradigm predicts that the clustering of dark matter halos depends not only on their mass but also on their formation time (the time at which the halo mass of the main progenitor has reached half of the final halo mass), an effect often called assembly bias.  Specifically, low-mass halos (M$_{halo}$ $<$ 6.15$\times$10$^{12}$ $h^{-1}$ M$_{\odot}$) that assemble early are much more strongly clustered than those that assemble late (see also Harker et al. 2006; Gao \& White 2007; Wechsler et al. 2006; Wetzel et al. 2007).  Clustering, measured using the 2-point autocorrelation function, increases with increasingly early formation times, and the signal is stronger as the length scale increases.  Fakhouri \& Ma (2009; 2010) find that the formation time of dark matter halos in the Millennium Simulation is earlier with increasing environmental density, measured using the dark matter mass within 7 Mpc of a galaxy.  They further find that mass growth proceeds differently in different environments, with more mergers in high-density regions and more diffuse accretion in low-density regions.  As Gao et al. (2005) point out, galaxy properties may depend on the assembly history of halos, and therefore models that ignore the age dependence of clustering do so at their peril.  

Berlind et al. (2006) examine SDSS groups and find that central galaxy color is correlated with clustering, but only for the most massive ($>$10$^{14}$ M$_{\odot}$) galaxy groups.  Groups with less red central galaxies cluster more than groups with redder central galaxies.  They conclude that massive halos that formed earlier contain redder galaxies than more recently-formed halos.  However, Yang et al. (2006) use the emission- and absorption-line strength in a galaxy's spectrum to study the SFR of the central galaxies of SDSS groups.  They find that group halos at all masses with central galaxies with lower SFRs are more clustered than groups with highly star-forming central galaxies.  Although Berlind et al. (2006) point out that emission- and absorption-line strength in a luminous red galaxy does not necessarily reflect its $g - r$ color, it is not clear why these two groups get these seemingly opposing results.

Much work has been performed studying assembly bias using SAMs or HODs.  For example, Gonzalez \& Padilla (2009) use a SAM to study the effect of the environment on galaxy SFRs.  They find that most of the differences in the SFR between galaxies in different large-scale environments can be explained by the environmental dependence of the halo mass function.  Further, they conclude that assembly bias is the most likely candidate driving the small differences in the populations beyond those explained by the halo mass function.  

Croton et al. (2007) come to a different conclusion using a SAM built into the Millennium Simulation.  They compare the two-point autocorrelation functions for the galaxies formed in their SAMs to a population that has been shuffled randomly into halos of the same mass (maintaining the relative separation between the central and its satellites).  Shuffling reduces the clustering of the entire sample.  They then change their shuffling scheme to also account for halo formation time (t$_{50\%}$) or concentration in addition to halo mass, but find that neither of these halo properties can account for the clustering differences seen in the SAM population versus the shuffled population.  They conclude that another unknown aspect of halo assembly or environment must be causing a large fraction of the bias.  

Zentner et al. (2014) find that assembly bias affects HOD modeling by comparing the HOD fits to mock catalogues with and without assembly bias.  They find that reasonable levels of assembly bias in the population can lead to systematic errors in the galaxy-halo connection inferred using standard HOD models, and conclude that incorporating assembly bias effects into future HODs should be a priority. 

Jung, Yee \& Li (2014) use a SAM to study whether the stellar mass of galaxies with the same halo mass is affected in regions of different large-scale (7 h$^{-1}$ Mpc) density.  They find a small difference, specifically that low-mass ($<$10$^{12}$ M$_{\odot}$) halos have slightly higher stellar masses in high-density environments.  

In this paper we compare halo mass growth, stellar mass assembly and the star formation histories of galaxies from $z$=6 to $z$=0 in a large-scale ($\sim$20 Mpc) overdensity and underdensity cosmologically simulated using the adaptive mesh refinement (AMR) Eulerian hydrodynamical code \textit{Enzo} (Bryan et al. 2014).  At their respective volumes, they represent +1.8$\sigma$ and -1.0$\sigma$ fluctuations.  These simulations provide us with an opportunity to compare different large-scale environments within the same fully hydrodynamical cosmological simulation performed at sub-kpc resolution.  It is important to compare across different environments within the same simulation, as all physical processes are modeled in the same way.

Our goal is to determine whether the stellar mass of galaxies is universally related to the halo mass of galaxies, or if it also depends on environment.  After a brief description of  our simulations (Section \ref{sec:method}), we discuss our galaxy selection technique and our method for determining quantities for each galaxy in Section \ref{sec:galaxyselection}.  In our results we compare the formation histories of galaxies in different environments, binned either by their final halo mass (Section \ref{sec:halo}) or stellar mass (Section \ref{sec:stars}).  In Section \ref{sec:discussion} we discuss several possible causes for the higher stellar masses found in galaxy halos in the overdense environment, ending with our comprehensive interpretation in Section \ref{sec:story}.  We then compare our results to other theoretical work on this issue (Section \ref{sec:comparison}).  Finally, in Section \ref{sec:conclusions} we summarize our conclusions and discuss the implications of our results.

\section{Method}\label{sec:method}

For the details of our simulations, we refer the reader to Cen (2012), although for completeness we reiterate the main points here.  We perform cosmological simulations with the AMR Eulerian hydrodynamical code \textit{Enzo} (Bryan 1999; Bryan et al. 2014).  We use cosmological parameters consistent with the WMAP7-normalized LCDM model (Komatsu et al. 2011):  $\Omega_M$ = 0.28, $\Omega_b$ = 0.046, $\Omega_\Lambda$ = 0.72, $\sigma_8$ = 0.82, $H_o$ = 100 $h$ km s$^{-1}$ Mpc$^{-1}$ = 70 km s$^{-1}$ Mpc$^{-1}$, and $n$ = 0.96.  We first ran a low resolution simulation with a periodic box of 120 $h^{-1}$ Mpc on a side, and identified two regions:  an overdensity centered on a cluster and an underdensity centered on a void at $z = 0$.  We then resimulated each of the two regions separately with high resolution, but embedded within the outer 120 $h^{-1}$ Mpc box to properly take into account large-scale tidal field effects and appropriate fluxes of matter, energy and momentum across the boundaries of the refined region.

The overdense refined region, or C box, is 21 $\times$ 24 $\times$ 20 $h^{-3}$ Mpc$^3$.  The central cluster has an M$_{200}$ of $\sim$2 $\times$ 10$^{14}$ M$_\odot$ with a virial radius (r$_{200}$) of 1.3 $h^{-1}$ Mpc.  The underdense refined region, or V box, is somewhat larger, at 31 $\times$ 31 $\times$ 35 $h^{-3}$ Mpc$^3$.  At their respective volumes, they represent +1.8$\sigma$ and -1.0$\sigma$ fluctuations where $\sigma$ is the density variance on the volume of the C and V boxes.  Although these are large-scale over- and underdense environments, these high-resolution boxes are much larger than the cluster or the void at their centers.  Thus, there are galaxies at a range of local densities in both boxes, and there is substantial overlap of local densities between the two volumes (Tonnesen \& Cen 2012).

In both refined boxes, the minimum cell size is 0.46 $h^{-1}$ kpc, using 11 refinement levels at $z = 0$.  The initial conditions for the refined regions have a mean interparticle separation of 117 $h^{-1}$ kpc comoving, and a dark matter particle mass of 1.07 $\times$ 10$^8$ $h^{-1}$ M$_\odot$.   

The simulations include a metagalactic UV background (Haardt \& Madau 1996), a model for shielding of UV radiation by neutral hydrogen (Cen et al. 2005), and metallicity-dependent radiative cooling (Cen et al. 1995).  The fraction and density of neutral hydrogen is directly computed within the simulations.  Star particles are created in gas cells that satisfy a set of criteria for star formation proposed by Cen \& Ostriker (1992), and reiterated with regards to this simulation in Cen (2012).  Briefly, A star particle is created if the gas
in a cell at any time meets the following three conditions simultaneously: (1) contracting flow, (2) cooling time less than
dynamical time, and (3) Jeans unstable.  A star particle of mass $m_*$ = $c_*m_{gas}{\Delta}t/t_*$ is created (the same
amount is removed from the gas mass in the cell), where ${\Delta}t$ is the time step, $t_*$ = max($t_{dyn}$,10$^7$ yr), $t_{dyn}$ = $\sqrt{3\pi/(32G\rho_{tot})}$ is the dynamical time of the cell, $m_{gas}$ is the baryonic gas mass in the cell, and $c_*$ = 0.03 is the star formation efficiency.  Each star particle has a mass of $\sim$10$^6$ M$_\odot$, which is similar to the mass of a coeval globular cluster.  Once formed, the stellar particle loses mass through gas recycling from Type II supernovae feedback, and about 30\% of the stellar particle mass is returned to the ISM within a time step.  Supernovae feedback is implemented as described in Cen (2012):  feedback energy and ejected metal-enriched mass are distributed into 27 local gas cells centered at the star particle in question, weighted by the specific volume
of each cell, which is to mimic the physical process of  supernova blastwave propagation that tends to channel energy, momentum, and mass into the least dense regions (with the least resistance and cooling). We allow the whole feedback process to be hydrodynamically coupled to surroundings and subject to
relevant physical processes, such as cooling and heating, as in
nature.  The simulation used in this paper has compared several galaxy properties that depend critically on the feedback method to observations and found strong agreement (Cen 2011a-c; Cen 2012).  

We do not include a prescription for AGN feedback in this simulation, and as a result, our simulation overproduces luminous galaxies at the centers of groups and clusters of galaxies.  This is discussed in detail in Cen (2011c), who shows that the luminosity function of the simulated galaxies agrees well with observations at $z$=0 except at the high-luminosity end.  When Cen (2011c) adds an AGN feedback correction in the post-simulation analysis that most strongly affects halos with masses greater than 10$^{13}$ M$_{\odot}$ or galaxies with stellar masses above 4$\times$10$^{12}$ M$_{\odot}$, the simulated luminosity function also agrees with observations at the high luminosity end.  We do not include any post-simulation AGN correction because of uncertainties in its implementation, particularly across a large redshift range (see discussion in Tonnesen \& Cen 2014), and as all of the galaxies we examine in this paper have halo masses less than 10$^{13}$ M$_{\odot}$, we would expect the correction to be minor in any case.

\subsection{Galaxies}\label{sec:galaxyselection}

We use HOP (Eisenstein \& Hut 1998) to identify galaxies using the stellar particles.   HOP uses a two-step procedure to identify individual galaxies. First, the algorithm assigns a density to each star particle based on the distribution of the surrounding particles and then hops from a particle to its densest nearby neighbor until a maximum is reached. All particles (with densities above a minimum threshold, $\delta_{outer}$) that reach the same maximum are identified as one coherent group.  The densest star particle is considered the center of the stellar group, or galaxy.  In the second step, groups are combined if the density at the saddle point which connects them is greater than $\delta_{saddle}$.  The minimum number of star particles in a group that HOP will return as a galaxy is 20, but this lower mass limit of 2$\times$10$^7$ M$_{\odot}$ is well below the lowest mass of the galaxies followed in this paper at $z$=0, which is well above 10$^9$ M$_{\odot}$.  We use HOP because of its physical basis, although we expect similar results would be found using a friends-of-friends halo finder.  HOP has been tested and is robust using reasonable ranges of values (e.g. Tonnesen, Bryan, \& van Gorkom 2007).

In this paper we measure the stellar mass (M$_*$), dark matter halo mass (M$_{\mathrm{halo}}$), and SFR of galaxies.  Stellar mass is determined by adding the mass of each star particle identified by HOP to belong to a galaxy.  The dark matter halo mass is calculated by summing the mass of all the dark matter particles out to r$_{200}$ (the radius within which the average density of the dark matter halo is 200 times the critical density).  

We also categorize our galaxies into centrals and satellites.  Central galaxies are those galaxies that are not within r$_{200}$ of any more massive galaxy, using the stellar mass as identified with HOP.  Satellite galaxies are defined as within r$_{200}$ of a more massive galaxy.  Our qualitative results do not depend on our specific choice of radius, and are qualitatively the same if we use a radius of 2 r$_{200}$.  In this paper, because we are studying the connection between halo mass and stellar mass, we only focus on central galaxies.

Lackner et al. (2012) describe in detail the merger trees that we will use in this paper, updated to include additional redshift outputs.  Briefly, the method tracks the stellar particles from one galaxy to another in sequential redshift slices.  If most of the stars in galaxy B at redshift 0.05 are in galaxy A at redshift 0, then we consider galaxy B to be the parent of galaxy A.  If several galaxies contribute stars to galaxy A, the galaxy that contributes the most stellar mass to galaxy A is the main progenitor.  A galaxy can be tracked to the minimum stellar mass identified by HOP, $\sim$2$\times$10$^7$ M$_{\odot}$, and this sets the minimum initial mass possible for galaxies we track.  We note that not all galaxies are initially identified with such a small set of particles.

In this paper we follow the stellar and dark matter mass histories of galaxies, so select only galaxies that have been tracked through the merger trees from a minimum redshift.  As we are examining whether galaxy stellar mass grows differently in low- versus high-density environments, we focus on galaxies that can be tracked to at least a redshift of 1.  We compared these results to a sample that could be tracked to at least a redshift of 3 to include the peak of the SFRD (Hopkins \& Beacom 2006; Karim et al. 2011; Seymour et al. 2008; Brinchmann et al. 2004; Reddy \& Steidel 2009; Bouwens et al. 2007), and found no qualitative differences in our results.  

In fact, we also find that if we compare the entire set of HOP-identified galaxies in each environment (as in Section \ref{sec:comparison}), we find the same results as when we focus on our tracked samples of galaxies.  This is an important verification exercise because demanding that a galaxy can be tracked to $z$=1 is essentially demanding that a galaxy has a stellar mass of at least 2$\times$10$^7$ M$_{\odot}$ by that redshift.  Particularly in our lowest mass bin, this will tend to select galaxies with higher $z$=0 stellar masses.  Because we find the same results when we include all galaxies at individual outputs, we know that this selection criterium does not effect our results.  However, as one of the goals of this paper is to consider the growth of galaxies over time, we focus on the tracked sample.

\section{Results}
\subsection{Matching Halo Mass}\label{sec:halo}

In order to compare the properties of galaxies over time, we have two galaxy samples: those in the large-scale underdensity (V box) and those in the large-scale overdensity (C box), both of which must be tracked to at least $z$=1.  Recall that in this paper we only consider central galaxies.  While this includes most galaxies in the V box (more than 90\% of galaxies at $z$=0), at lower masses in the C box there are a large fraction of satellite galaxies ($\sim$35\% at $z$=0), so the mass growth of the galaxies we follow is not necessarily representative of the C box population as a whole.  In order to present our results the most clearly, we have binned both samples by halo mass at $z$=0.  We have verified that the specific mass ranges do not affect our results:  we have selected our samples in several different ways and found that our results are robust, even when we use C box samples that have slightly lower halo masses than their V box counterparts.  We consider four mass bins, with M$_{halo}$/M$_{\odot}$ between 10$^{11}$-10$^{11.5}$, 10$^{11.5}$-10$^{11.9}$, 10$^{11.9}$-10$^{12.3}$, and 10$^{12.3}$-10$^{12.9}$.  

\begin{figure}
\includegraphics[scale=0.96,trim= 0mm 0mm 0mm 0mm, clip]{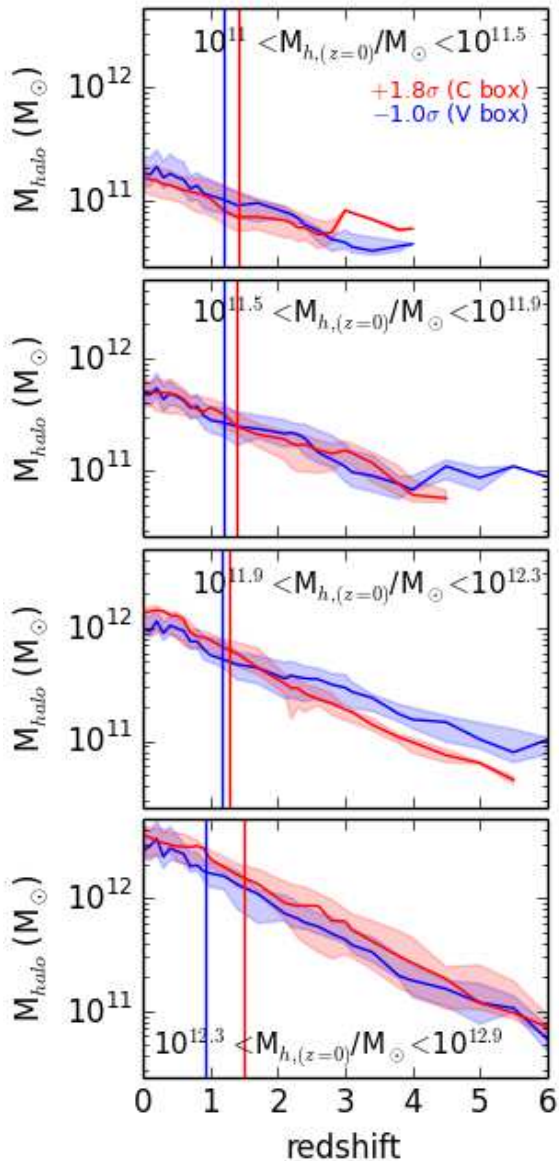}
\caption{The median halo mass of tracked galaxies from a possible maximum redshift of six.  Galaxies are binned according to halo mass at $z$=0.  The solid lines are the median value of the sample of tracked galaxies, and the shading denotes the middle range of values (25-75 percentiles).  The V box galaxies (underdense environment) are shown in blue and the C box galaxies (overdense environment) are shown in red.  The vertical lines are the median redshift at which the halo has grown to half of its final mass.  Our galaxy sample selection has selected galaxies with similar $z$=0 halo masses in the V and C boxes.  See Section \ref{sec:halo} for details and Section \ref{sec:halohalo} for discussion.}\label{fig-hmbhm}
\end{figure}

\begin{figure}
\includegraphics[scale=0.96,trim= 0mm 0mm 0mm 0mm, clip]{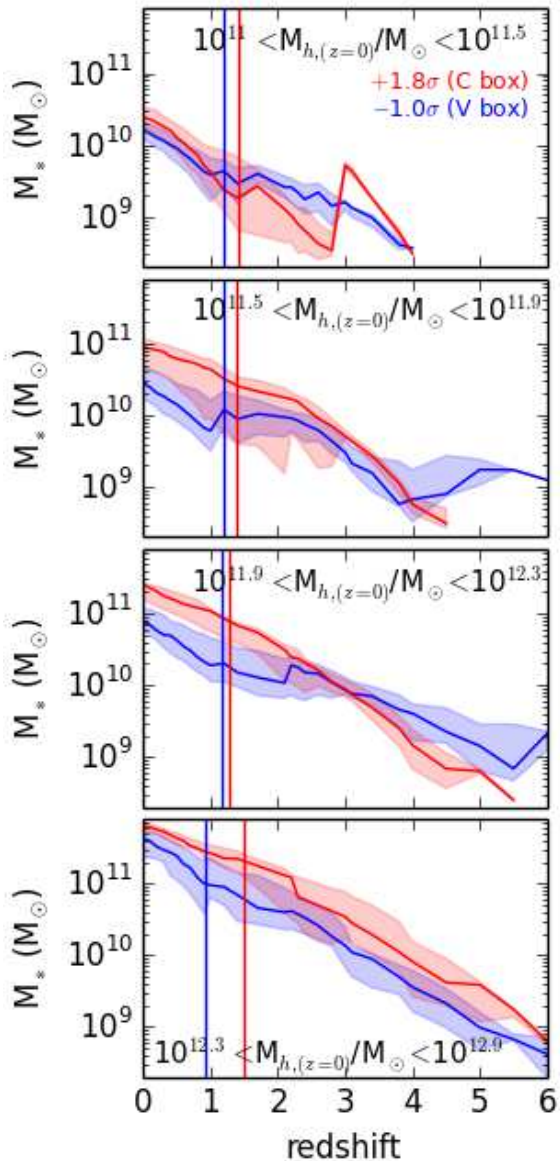}
\caption{The median stellar mass of tracked galaxies binned by halo mass at redshift 0.  Lines are as in Figure \ref{fig-hmbhm}.  The C box galaxies in the overdense environment (red) tend to have higher final stellar masses than the V box galaxies in the underdense environment (blue), and the crossover redshift from lower to higher stellar masses increases with increasing redshift.  See Section \ref{sec:halostars} for discussion.}\label{fig-hmbsm}
\end{figure}

\begin{figure}
\includegraphics[scale=0.96,trim= 0mm 0mm 0mm 0mm, clip]{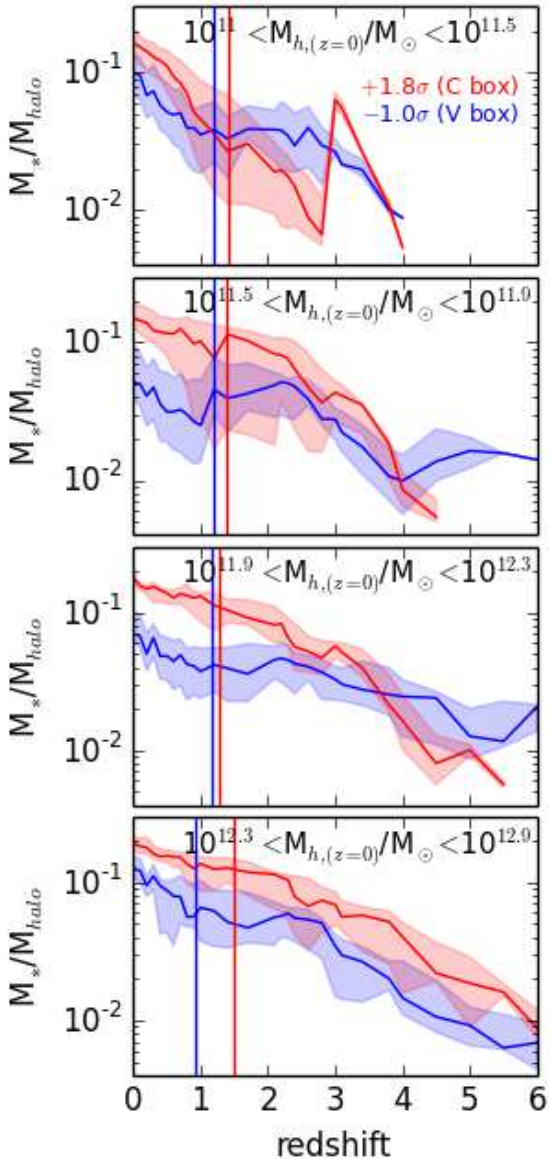}
\caption{The median stellar mass to halo mass ratio (SMHM) of tracked galaxies binned by halo mass at redshift 0.   Lines are as in Figure \ref{fig-hmbhm}.  The galaxies in the overdensity (C box, red lines) have higher SMHM ratios than those in the underdensity (V box, blue lines).  This result holds even comparing between galaxies in different mass bins.  See Section \ref{sec:halostars} for discussion.}\label{fig-hmbsmhmr}
\end{figure}

\begin{figure}
\includegraphics[scale=0.96,trim= 0mm 0mm 0mm 0mm, clip]{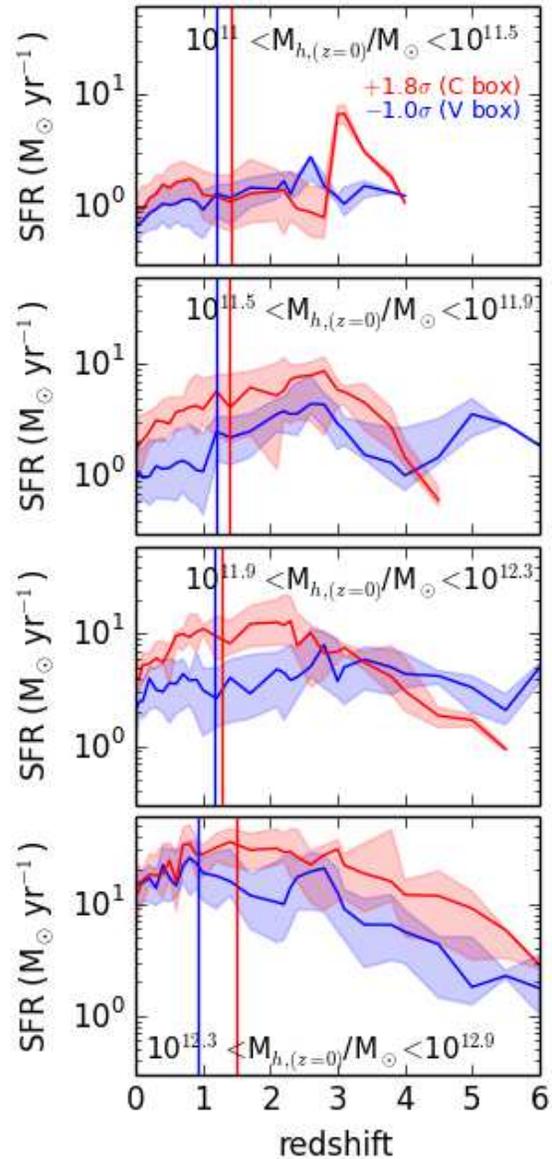}
\caption{The median SFR of tracked galaxies binned by halo mass at redshift 0.  Lines are as in Figure \ref{fig-hmbhm}.  The SFRs of galaxies in the overdensity (C box, red lines) tend to be lower than those of galaxies in the underdensity (V box, blue lines) at $z$$\ge$4, but become higher at lower redshifts.  See Section \ref{sec:halosfr} for discussion.}\label{fig-hmbsfr}
\end{figure}

\begin{figure}
\includegraphics[scale=0.96,trim= 0mm 0mm 0mm 0mm, clip]{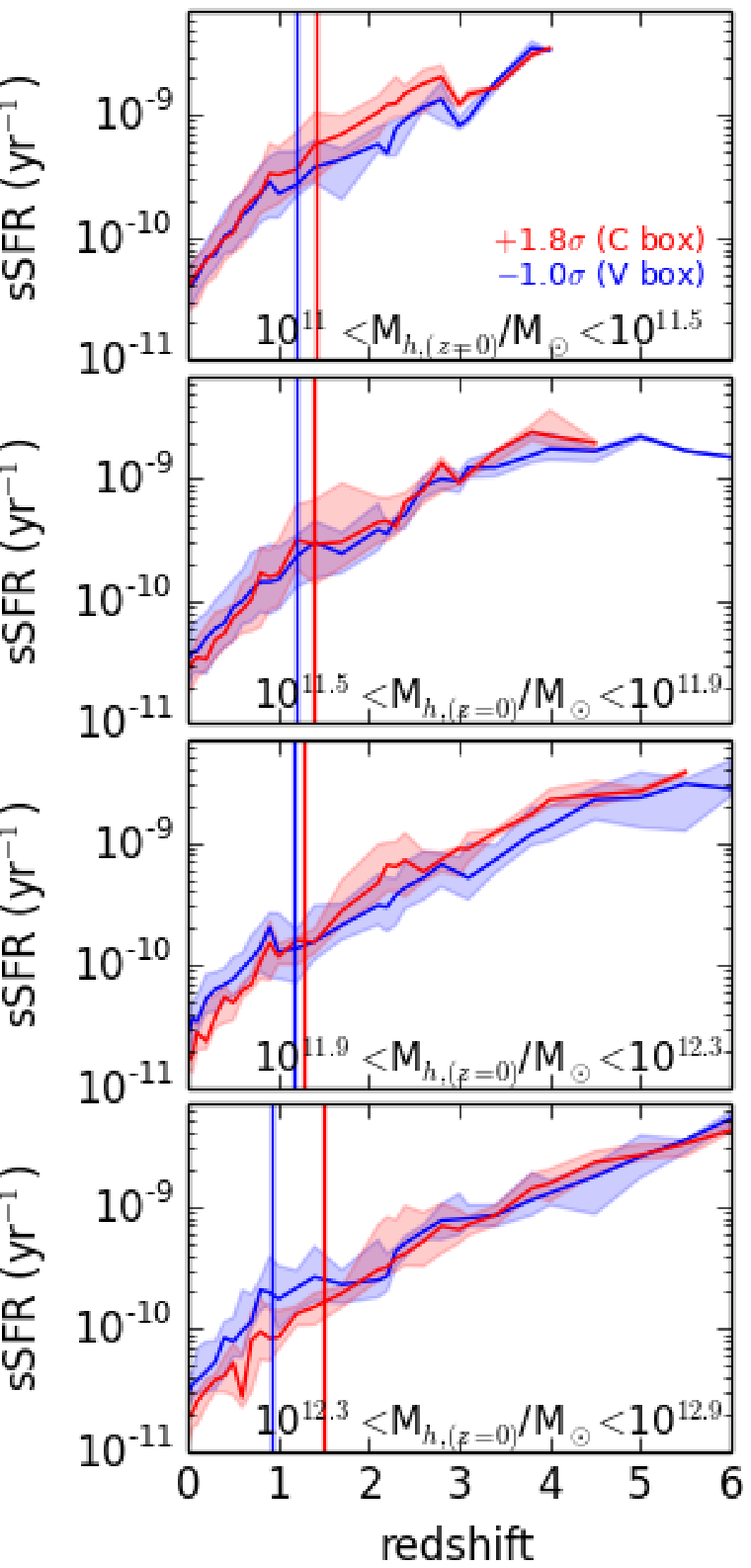}
\caption{The median sSFR of tracked galaxies binned by halo mass at redshift 0.  Lines are as in Figure \ref{fig-hmbhm}.  Galaxies follow the sSFR-stellar mass relation.  Comparing galaxies with the same $z$=0 halo mass, we find that at early times the sSFR of galaxies in the overdensity (C box, red lines) is higher with respect to the sSFR of galaxies in the underdensity (V box, blue lines) than at $z$=0.  See Section \ref{sec:halosfr} for discussion.}\label{fig-hmbssfr}
\end{figure}

In Figures \ref{fig-hmbhm}-\ref{fig-hmbssfr} we show several properties of our galaxies over our tracked redshift range.  In all figures, the solid lines are the median value of the sample, and the shading denotes the middle range of values (25-75 percentiles).  The V box galaxies (underdense environment) are shown in blue and the C box galaxies (overdense environment) are shown in red.  The vertical lines are the median redshift at which the halo has grown to half of its final mass (formation time).  Until our minimum tracking redshift, $z$=1, new galaxies can be added to the set, which may decrease the median halo or stellar mass values.  This can result in the sometimes dramatic differences in median values as galaxies are added to the sample, particularly in the lowest mass bin.  These jumps are not important to our results, which as we have discussed are insensitive to our sample selection, including the redshift to which we track galaxies.  

In order to speed post-processing of our galaxy parameters, we use coarse radial bins to determine r$_{200}$, and do not calculate the average overdensity continuously.  This means that the halo mass can vary, as seen particularly at late times in the underdense box, because a small difference in the radial density profile can result in a large variation in our r$_{200}$ due to the bin size, and therefore M$_{halo}$ may vary.  These variations also do not affect our results.  

%fig 1

%fig 2

%fig 3

%fig 4

%fig 5

\subsubsection{Halo Mass}\label{sec:halohalo}

First we consider the halo mass of our galaxy samples in Figure \ref{fig-hmbhm}.  We can compare the formation time (or halo age) of our tracked galaxies in the overdense and underdense environments.  As in Gao et al. (2005), we define the age of a halo using the redshift at which the mass of the main progenitor halo is 50\% of the final halo mass.  Our outputs are in redshift steps of 0.05 and 0.1 in the C and V box, respectively, which sets our uncertainty on the halo age of any individual galaxy.  If we do not start tracking a galaxy until after it has reached half of its final mass, we set the formation time to the earliest time tracked.  We compare our median formation time including these estimated values to a median formation time calculated using only the formation times for galaxies that we begin tracking when they are less than 50\% of their final halo mass, and find that for the three highest bins the change is minuscule.  The lowest bin, that tends to begin tracking galaxies at later times, shifts to earlier median formation times when only the smaller sample is used.  In the lowest mass bin, from 10$^{11}$-10$^{11.5}$ M$_{\odot}$, the formation times go from being similar in the two environments to being at higher redshifts in the overdense sample. 

If we compare across different masses in the same environment, we find that in our underdense samples, as in Behroozi et al. (2013), the halo age decreases as halo mass increases.  This is evident when comparing the highest mass bin to any of the lower mass bins.  In contrast, the overdense galaxy samples tend to have slightly decreasing formation times with increasing mass until the most massive halo bin (10$^{12.3}$-10$^{12.9}$ M$_{\odot}$), which has the oldest formation time.  It is possible that our selection criterion that a galaxy must remain a central galaxy to $z$=0 eliminates the older low-mass galaxies, which will effect the median formation time in the overdense environment more strongly because a higher fraction of galaxies are satellites.

Comparing between galaxies of the same halo mass in underdense versus overdense environments, we find that in general halos in the overdense region have older or similar formation times than galaxies in the underdense environment.  This tendency agrees with the halo bias findings of Gao et al. (2005), that more clustered halos tend to collapse first.  As we expect, the biggest difference is in the highest mass bin (10$^{12.3}$-10$^{12.9}$ M$_{\odot}$), where we are the least likely to exclude halos from our sample because they become satellites.   

We find that, in agreement with the similar halo ages, the halo growth curves are very similar in the lower mass bins (10$^{11}$-10$^{11.5}$ M$_{\odot}$ and 10$^{11.5}$-10$^{11.9}$ M$_{\odot}$), particularly after about $z$=2.5.  However, in the highest M$_{halo}$ bin, the C box galaxies' halo growth flattens from about $z$=1, while the galaxies in the V box continue to grow more massive.  

In summary, in agreement with previous N-body simulations (e.g. Gao et al 2005 and see our Introduction), we find evidence of assembly bias.  Also, in the highest mass bin (10$^{12.3}$-10$^{12.9}$ M$_{\odot}$), we see a hint that at low redshift ($z$$<$1) the dark matter mass growth of massive halos flattens in galaxies in the overdense environments in comparison to halos in the underdense environment.  

\subsubsection{Stellar Mass}\label{sec:halostars}

In Figure \ref{fig-hmbsm} we focus on the stellar mass of our galaxies.  In the overdense (C box) environment, the stellar mass growth from $z$=1 to $z$=0 is the steepest in the lowest halo mass bin (10$^{11}$-10$^{11.5}$ M$_{\odot}$), indicating a trend of downsizing in stellar mass growth with decreasing redshift.  In all mass bins in the underdense environment and in the upper three mass bins in the overdense sample the stellar mass growth from $z$$\le$1 is similar.  However, the galaxies in the underdense environment grow more than the galaxies in the overdense environment in all but the lowest mass bin, indicating that stellar growth shifts from overdense to underdense regions over time.  

In all of the halo mass bins, the median stellar mass of galaxies in the overdense box is higher than that of galaxies in the underdense box by $z$=0.  In all but the highest mass bin (10$^{12.3}$-10$^{12.9}$ M$_{\odot}$), the median stellar mass in the C box galaxies begins lower than the median stellar mass for the V box galaxies.  The redshift at which the C box stellar mass crosses the V box stellar mass tends to decrease with decreasing mass--from the highest mass bin always having higher stellar mass galaxies in the C box to a crossover at $z$$\sim$1 in the lowest mass bin.  However, the stellar mass growth at very low redshift ($z$$\le$1) is generally flatter in the C box galaxies than in the V box galaxies, indicating less recent star formation, in agreement with observations (e.g. Szomoru et al. 1996).  

The stellar mass to halo mass (SMHM) ratio, shown in Figure \ref{fig-hmbsmhmr}, also reflects the higher stellar mass in halos in the overdense environment.  In general, even when comparing across different halo mass bins, we see that the SMHM ratio is higher in the overdense environment, showing that the halo mass distribution does not effect this result.  In other words, by $z$$=$0, the SMHM ratio of overdense galaxies in any mass bin is higher than the SMHM ratio of underdense galaxies in any halo mass bin.  

Comparing within each environment, we find that the SMHM ratio at $z$=0 is the highest for galaxies with the highest halo masses, unlike previous results (e.g. Moster et al. 2012, Guo et al. 2010, Behroozi et al. 2013 and references therein) that the peak SMHM ratio will be found for galaxies with halo masses of about 10$^{12}$ M$_{\odot}$.  These previous results use several methods to match results to existing data, and we do not call their overarching SMHM results, that do not consider different large-scale environments, into question.  Indeed, our SMHM ratio tends to be too high.  This may indicate that our feedback mechanisms are inefficient, particularly at the lowest and highest masses, which we will discuss in detail below in Section \ref{sec:feedback}.  We conclude that the lowest mass and highest mass halos have the most cold gas that is able to form stars, either by never shock-heating accreting gas to high temperature or by having a high enough central density to cool gas and form stars.   As our feedback prescription is the same across all environments, the differences between galaxies in the overdense and underdense regions should be real and not due to our particular numerical methodology, although we cannot eliminate the possibility that our specific feedback scheme could smooth out or exacerbate any differences between galaxies in different environments.

We conclude that the SMHM ratio differs in these differing large-scale environments, with central galaxies in overdense environments having higher SMHM ratios than those of central galaxies in underdense environments.  

\subsubsection{Star Formation Rate}\label{sec:halosfr}

We next focus on the SFRs of the galaxies in Figure \ref{fig-hmbsfr}.  For all galaxy samples but the highest mass galaxies in the underdense environment, the SFR initially increases to an early peak, then decreases to $z$$=$0.  This follows our expectations, because as shown in Cen (2011c) our simulated SFR density history agrees well with observations that show a peak between $z$=2-3 followed by a declining SFRD.  

When we compare galaxies of the same mass in different environments, we find that for all but the highest mass galaxies, the SFR in galaxies in the overdense environment begins lower than the SFR in galaxies in the underdense environment.  The SFR of the C box galaxies increases with respect to the SFR of the V box galaxies as the universe ages, eventually overtaking the SFR of V box galaxies.  The redshift at which the C box SFR becomes higher than the V box SFR tends to decrease for decreasing halo mass.  In our highest mass bin (10$^{12.3}$-10$^{12.9}$ M$_{\odot}$), the C box SFR begins higher than the V box SFR, but decreases quickly after $z$=1.  

In Figure \ref{fig-hmbssfr} we see that in either environment at $z$=0 our galaxies follow the sSFR-stellar mass relation, with lower mass galaxies having higher sSFRs.  We also see that at early times the sSFR of galaxies in the overdense environment tends to be somewhat higher than the sSFR of galaxies in the underdense environment.  At late times, the $z$=0 sSFR is either similar in the two environments, or higher in the V box than in the C box, probably reflecting the sSFR-M$_*$ relation.  This may also agree with the Szomoru et al. (1996) observational result that galaxies in the Bootes Void tend to lie above the Tully-Fisher relation using the B-band magnitude.  
  
The comparison between the star formation histories of these galaxies align with the comparative stellar mass growth of these galaxies.    

\subsection{Matching Stellar Mass}\label{sec:stars}

In order to compare the properties of galaxies using a more easily observed quantity, we verify our results when we bin galaxies by their stellar mass at $z$=0.  We find that, as when we bin by halo mass, galaxies in the overdense region tend to have higher SMHM ratios.  Unlike when we bin by halo mass, we find that the SFRs and sSFRs in the two environments are more similar by $z$=0, as we might expect given that the final stellar mass is quite similar.  However, the $z$=0 SFR and sSFR of our massive galaxies in the underdense region tends to be slightly higher than in the overdense region.  Thus, depending on the stellar masses of the galaxies we compare, the sSFR is the same across different environments, or galaxies of the same mass are bluer in underdense environments than in overdense environments (Yang et al. 2006; Rojas et al. 2004; Grogin \& Geller 1999,2000; Szomoru et al. 1996).  

\section{Discussion}\label{sec:discussion}

We will now discuss several possible causes of this difference in the SMHM ratio in galaxies in different environments.  Finally, we will propose our synthesized picture of how higher-density large-scale environments could produce galaxies with high SMHM ratios.

\subsection{Formation Time}\label{sec:t50}

We first discuss whether the formation time, defined as in Gao et al. (2005) and marked with vertical lines in Figures \ref{fig-hmbhm}-\ref{fig-hmbssfr}, is related to the increased SMHM ratios in the overdense environment.  We see a clear trend that the C box galaxy populations tend to have earlier formation times and higher SMHM ratios.  Although the sample used in this paper of overdense galaxies is slightly more massive than the galaxies in the underdense region, we have verified that this result does not depend on the mass distribution of the two samples by selecting an overdense galaxy sample that has slightly lower masses than the underdense sample.  

However, we find that more subtle differences in formation time are not clearly related to the final SMHM ratios.  Examining Figure \ref{fig-hmbsmhmr}, we see that the two largest differences in the SMHM ratios are in the halo mass bins from 10$^{11.5}$ - 10$^{11.9}$ and 10$^{11.9}$ - 10$^{12.3}$.  These mass bins have the smallest differences in the formation times between the overdense and underdense populations.  The largest difference in formation times is in the highest mass bin (10$^{12.3}$-10$^{12.9}$ M$_{\odot}$), and the difference between the final SMHM ratios in the two environments is among the smallest.
 
To summarize, we find that galaxies in the overdense environment have both earlier formation times and higher SMHM ratios than galaxies in the underdense environment.  However, an earlier formation time does not necessarily lead to a higher SMHM ratio, as seen by comparing the galaxies in the underdense environment in Figure \ref{fig-hmbsmhmr}, and, as we discussed above, a larger difference in formation times does not force a larger difference in SMHM ratio.  

\subsection{Tidal Stripping of DM Halos?}\label{sec:tides}

Several recent works have argued that assembly bias, the earlier formation times of halos in higher density environments, can be explained by the increased tidal interactions with neighbors in higher density environments.  N-body simulations have shown that at late times, the tidal field from massive neighbors can halt halo growth and even in some cases reduce halo mass (Avila-Reese et al. 2005; Maulbetsch et al. 2007; Wang et al. 2006; Diemand et al. 2007; Hahn et al. 2009; McBride et al. 2009; Wang et al. 2011).  Wang et al. (2007) specify that halos in overdense environments cannot accrete dark matter because of the high velocity dispersion of the dark matter.  These works generally find that tidal effects are strongest for low-mass halos. 

We might expect to see several signatures of this effect on our results.  First, we would expect slower late-time growth of halos in the overdense region compared to halos in the underdense region.  Secondly, we would expect that within the overdense box, low mass halos would have slower late-time growth than high mass halos.  Finally, we would expect tidal effects to cause the largest difference between galaxies in the overdense versus underdense environment for galaxies with the lowest halo mass.

In agreement with our first prediction, we see that the highest halo mass bin (10$^{12.3}$-10$^{12.9}$ M$_{\odot}$) shows slower growth in the M$_{halo}$ of the galaxies in overdense regions at late times ($z$$<$1; Figure \ref{fig-hmbhm}).  Also, as discussed above (Section \ref{sec:t50}), we also find that the formation time of galaxies in the higher density environment tends to be earlier than in the lower density environment.  

However, when we test the second prediction by comparing different mass ranges within the overdense box, we do not find the expected trend.  As discussed in Section \ref{sec:halohalo}, the halo age is the oldest for the highest halo mass bin in the overdense box, indicating that the late-time halo growth of low mass halos is larger than that of the highest mass halos.  If we focus on halo growth from $z$=1 to $z$=0, we find that most mass bins increase by about the same factor.  A caveat to this result is that the smaller-scale environment within the large-scale overdensity is not necessarily the same surrounding galaxies of different masses.  In fact, the number of galaxies within 2 Mpc of the tracked galaxies decreases with decreasing halo mass, so a careful comparison may show that in the same smaller-scale density field, lower mass halos grow more slowly at late times.  

%fig 6
\begin{figure*}
\begin{center}
\includegraphics[scale=0.96,trim= 0mm 0mm 0mm 0mm, clip]{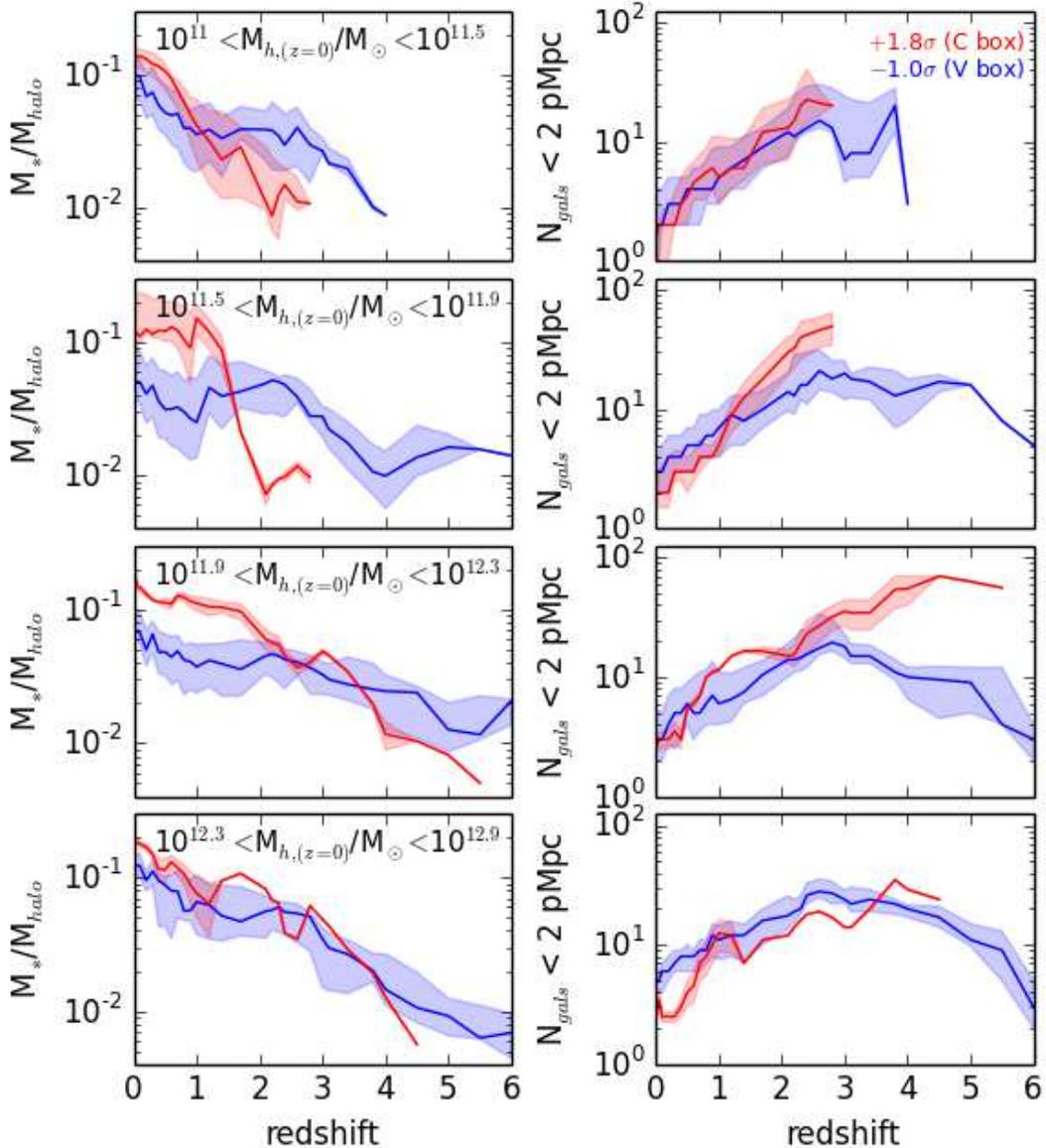}
\end{center}
\caption{The SMHM ratio (left) and the number of galaxies within 2 pMpc of galaxies (number of nearby neighbors plus the galaxy itself), with an additional criterium for the overdense (C box) sample that the $z$=0 number of nearby galaxies is less than or equal to 2 (3 including the galaxy).  Even if we select C box galaxies with lower local densities than that of V box galaxies, the SMHM ratio remains higher.  See Section \ref{sec:n1mpc} for discussion.}\label{fig-neighbors}
\end{figure*}

Finally, against our third expectation, the difference between the stellar mass of the lowest mass galaxies (10$^{11}$-10$^{11.5}$ M$_{\odot}$) is clearly the smallest of any of the mass bins, and the similarity of the SMHM ratio supports that finding.  Higher mass bins (10$^{11.5}$-10$^{11.9}$  M$_{\odot}$ and 10$^{11.9}$-10$^{12.3}$ M$_{\odot}$) have the largest SMHM ratio difference.  

Therefore, while we find evidence that tidal truncation of accretion and/or tidal stripping of dark matter halos is occurring in the overdense environment by comparing the late-time growth of halos, because the difference in formation time is not the largest for low-mass galaxies and because the difference in the $z$$=$0 SMHM ratio is the smallest for low-mass galaxies, we do not think that tidal effects are the main driver of our results.  

\subsection{Local Galaxy Density}\label{sec:n1mpc}

Tidal effects on galaxies may come from local halos, and we have not normalized for the local galaxy density.  Indeed, in all of the mass bins, the galaxies in the overdensity (C box) have a larger median number of galaxies within 2 physical Mpc (pMpc) of the tracked sample than the galaxies in the underdensity (V box).  In order to determine if the local galaxy density within 2 pMpc is in fact causing the differences in the SMHM ratios, we select a subset of C box galaxies in each mass bin that fit the additional criteria that at $z$=0 the number of galaxies within 2 pMpc of the C box galaxy is three or less, including the galaxy itself, meaning two or fewer neighbors.  We choose the upper limit of two neighbors because the median number of neighbors in the V box sample is two in our middle two mass bins (10$^{11.5}$-10$^{11.9}$ M$_{\odot}$ and 10$^{11.9}$-10$^{12.3}$ M$_{\odot}$), one in our lowest mass bin, and four in our highest mass bin.  Thus we can compare this "low local-density" C box sample to the complete, unchanged, V box sample.

In Figure \ref{fig-neighbors} we show the SMHM ratio and number of galaxies within 2 pMpc in this much reduced sample.  The small sample size of C box galaxies is most dramatic in the highest halo mass bins.  In the highest mass bin (10$^{12.3}$-10$^{12.9}$ M$_{\odot}$) there is only one galaxy in the C box sample until $z$=1.2, and only two galaxies thereafter.  The second highest bin also only has two galaxies in the C box (10$^{11.9}$-10$^{12.3}$ M$_{\odot}$), and the third bin has three (10$^{11.5}$-10$^{11.9}$ M$_{\odot}$).  The C box samples in the lowest mass bin, 10$^{11}$-10$^{11.5}$ M$_{\odot}$, has 22 galaxies.  Due to these small samples, the conclusions we can draw are necessarily tentative.  

While at early times the C box galaxies always have more neighbors, by $z$=1 the numbers are similar in the two environments, and by $z$=0 the number of neighbors in the underdense environment is generally slightly more than in the overdense environment (by construction, due to our selection criteria).  We note that this may indicate more mergers in the higher density environment, in agreement with Fakouri \& Ma (2009).  However, even using this specifically selected sample, the SMHM ratio is higher in the overdense (C box) sample.  Therefore, when examining a z$<$1 galaxy sample, simply forcing the number of galaxies to be the same in 2 Mpc regions is not enough to damp out larger-scale environmental effects.

\subsection{Satellites}\label{sec:satellites}

We next briefly consider whether satellites could have an effect on the SMHM ratio.  First, satellites will add to the total dark matter mass within one virial radius before the stellar particles merge to be identified as a single object by HOP, so more satellites might lead to lower SMHM ratios.  It is also possible that gravitational effects from an orbiting satellite could drive gas towards the center of the central galaxy and increase the SFR, resulting in higher SMHM ratios.  

However, the differences in the number of satellites between the overdense and underdense samples are quite small.  None of the central galaxies in the lowest mass bin have satellites identified by HOP (which identifies stellar groups of 20 particles, or about 2$\times$10$^7$ M$_{\odot}$).  In the second lowest halo mass bin, the median number of satellites is always zero, although the 75 percentile can extend to one satellite in both the C and V box samples.  In the second highest mass bin the median number of satellites varies between 0 and 1, with the overdense sample tending to have more.  Finally, in the highest mass bin, after $z$$\sim$3 the galaxies in the underdense sample have a median number of 1-2 satellites while the galaxies in the overdense sample have a median number of satellites between 0-1.5 (a median of 1.5 can occur with an even number of galaxies in the sample). 

While satellites may have something to do with the lower SMHM ratio in the underdense galaxies in the highest mass bin, it is unlikely that they play a role for the majority of galaxies.  

\subsection{Gas supply affected by the large-scale environment?}\label{sec:gassupply}

We now discuss whether the gas supply available to form stars in galaxies of the same halo mass may differ in different environments.  Clearly, the dark matter density in the overdense region is higher than in the underdense region.  If baryons generally trace dark matter, there must be therefore more gas available in the overdense region.  Indeed this is the case in our simulation, as the global baryon fraction differs by less than 5\% in the C and V boxes.  The question is then: why does more gas form stars in halos in the overdense region?  

In recent work, Cen (2014) has found that the number of cold streams feeding galaxies correlates with the SFR.  Therefore, we posit that if galaxies in the overdense environment have more cold streams, then more gas may be able to form stars.  Cen (2014) also argues that streams have higher densities in environments that are effected by gravitational heating.  In our higher density environment, individual galaxy halos are more likely to collapse within denser filaments or Zeldovich pancakes, which may result in more higher density cold streams.  

However, Fakhouri \& Ma (2010), examining the dark matter-only Millennium Simulation, find that although there is more diffuse dark matter around halos in high-density environments, the accretion rate of diffuse dark matter is higher in low-density environments.  They propose that this may be because diffuse material is dynamically hotter in high-density environments, and therefore unable to accrete onto halos.  Although determining the environmental dependence of galaxy accretion from cold streams is beyond the scope of this paper, in future work we will compare the cold streams accreting onto galaxies in overdense versus underdense regions.  

\subsection{Stronger Feedback}\label{sec:feedback}

We can now discuss the possible effects of including AGN feedback or changing the strength of SN feedback on our results.  

Including AGN feedback would have the most effect on high-mass galaxies (M$_{halo}$ $\ge$10$^{13}$ M$_{\odot}$), which are not the focus of this work.  While there would be some effect on the galaxies in our highest mass bin, it would be quite small (as these halos are $<$10$^{13}$ M$_{\odot}$).  Also, this effect would be smaller at earlier times, before the AGN became massive, although Schaye et al. (2010) find that AGN feedback begins affecting the SFR of massive galaxies as early as $z$$\sim$3.  However, we note that the stellar mass of the C box galaxies in the highest halo mass bin is always (to $z$$\sim$6, the entire time that galaxies are tracked in the simulation) higher than the stellar mass of the V box galaxies.  Therefore, we conclude that including AGN feedback would not strongly affect our results. 

As discussed above, we posit that much of the difference in the stellar mass of galaxies in underdense versus overdense environments is due to differing gas supply.  Therefore, decreasing the energy input from supernovae feedback would not bring the stellar masses of galaxies in these different environments into agreement and might increase the difference.  On the other hand, increasing the strength of SN feedback may bring the stellar masses of galaxies in these different environments into better agreement.  As shown in Figure 12 of Cen (2011c), the specific cold gas inflow rate is always higher than the sSFR, in both the C and V boxes, which is likely due to SN feedback regulating star formation rates.  Therefore, it may be possible to eliminate any differences with environment if we were to increase the strength of our feedback to lower the SFR at all halo masses.  As increasing the feedback strength would affect halos in both the overdense and underdense environments, we expect that a small change would not be enough to bring the stellar masses into agreement.  Therefore we are hesitant to invoke increased feedback to remedy this difference, as in several papers we have compared our simulations to observations and found good agreement.  For example, the SFRD in our two extreme regions brackets the observed global SFRD (Cen 2011c), and dramatically affecting the SFR of galaxies in our simulations would have a significant effect on this measurement.

As shown in Cen (2011c), the specific cold gas inflow rate is higher than the sSFR for galaxies in both the C and V boxes, indicating self-regulation of SF.  However, the difference between the specific inflow rate and the sSFR is highest at high redshift (z$>$3), and at these high redshifts V galaxies have a higher inflow to sSFR ratio.  Therefore, we find that SN feedback is making SF less efficient in galaxies in the underdense environment.  Because our current feedback scheme seems to affect the gas consumption rate in V galaxies more strongly than in C galaxies, we infer that increasing the feedback strength would only exacerbate the difference in the populations.

Although we cannot use SAMs to make strong predictions for hydrodynamical simulations, we note that in Jung et al. (2014) the authors vary their star formation feedback strength by a factor of four and find no qualitative difference in their result for low mass halos that galaxies in dense environments have higher stellar masses than galaxies in underdense environments.  

Finally, even if we devised a feedback prescription that eliminated the differences in SFR and therefore stellar mass in halos in different large-scale environments, our results are interesting because they indicate that gas accretion differs across different large-scale environments, and is not solely dependent on halo mass.  

\subsection{How environment affects the SMHM ratio}\label{sec:story}

We conclude that the SMHM ratio depends on more than the halo mass or local (2 pMpc or less) galaxy density at $z$=0.  Our results point to a picture in which the combination of earlier formation times, enhanced interactions, and more filaments are the likely causes of the increased SMHM ratios in galaxies in higher density environments.  

First, formation time plays a role.  Galaxies in the overdense region have earlier formation times.  An earlier formation time means that the galaxies in the overdense region grow more quickly at earlier times, when all galaxies have higher SFRs and more star-forming gas available.  Therefore, their higher halo mass at z$>$1 allows them to form more stars when rates are high.  As we are considering halos that all have $z$=0 masses below 10$^{13}$ M$_{\odot}$, the halos are not massive and hot enough to quench cold accretion and/or star formation at early times.  

Second, interactions with nearby neighbors play a role.  Galaxies in overdense environments have more nearby neighbors, particularly at early times.  In Figure \ref{fig-neighbors}, where we have forced the $z$=0 number of neighbors in the C box to be less than in the V box, we see that at early times the SMHM ratio is lower and the number of neighbors is higher in the galaxies in the overdense region than in those in the underdense region.  The SMHM ratio increases in the C box galaxies relative to the V box galaxies, as the number of neighbors decreases in the C box to equal the number of neighbors are V box galaxies.  As discussed in Cen \& Safarzadeh (2014), interactions first drive enhanced star formation in massive galaxies, then at later times affect lower mass galaxies.  In agreement with this scenario, in Figure \ref{fig-neighbors} (and Figure \ref{fig-hmbsmhmr}) the SMHM ratio of the C box galaxies crosses that of the V box galaxies at later times for lower mass galaxies.   

Finally, filamentary structure affects galaxy growth.  As we discuss in Section \ref{sec:gassupply}, Cen (2014) finds that number of filaments correlates with SFR.  Overdense regions are likely to have more filaments that can feed galaxies through smooth accretion and increase SFRs.  In particular, the growth in both the stellar and dark matter mass of galaxies in the overdense region happens at earlier times, when filamentary accretion is more likely to be able to strongly increase the mass of galaxies, and gas consumption into stars is more efficient  (Cen 2011c).  At later times, filamentary accretion becomes less important in general, which is why filaments can impact the SMHM ratio even of galaxies in the large scale overdensity that are in local underdensities by $z$=0.

The relative SMHM ratios of galaxies in different environments depends on the mass of the galaxies.  In the two most massive bins in Figure \ref{fig-hmbsmhmr} the difference between the SMHM ratio in the two environments peaks before $z$=0.  This is because gas accretion and star formation have slowed in the high mass galaxies in the overdense environment, but are still continuing in the high mass galaxies in the underdense environment.  The difference in the SMHM ratio remains similar from $z$=1 to $z$=0 in the second mass bin (10$^{11.5}$-10$^{11.9}$ M$_{\odot}$), and is increasing for galaxies in the lowest mass bin.  We see this trend with mass reflected in Figure \ref{fig-hmbssfr}, that the sSFR in the highest mass bin is lower in the overdense than in the underdense environment, while at the lowest halo masses the sSFR is similar in the overdense environment.  Therefore, we suspect that for even higher masses, by $z$=0 the difference in SMHM ratio would be entirely wiped out by the higher sSFR in the low density environment.  

In summary, the environment affects the formation time of halos, which in turn affects how gas-rich the universe is when they are massive enough to quickly accrete gas and form stars.  Also, galaxies in higher density large-scale environments have a larger number of neighbors, particularly at early times.  These interactions can drive gas instabilities and star formation.  Finally, galaxies in higher density environments are likely fed by more, and higher density, filaments. 

\section{Comparison with Other Work}\label{sec:comparison}

%fig 7
\begin{figure*}
%\begin{center}
\includegraphics[scale=0.79,trim=0mm 0mm 0mm 0mm, clip]{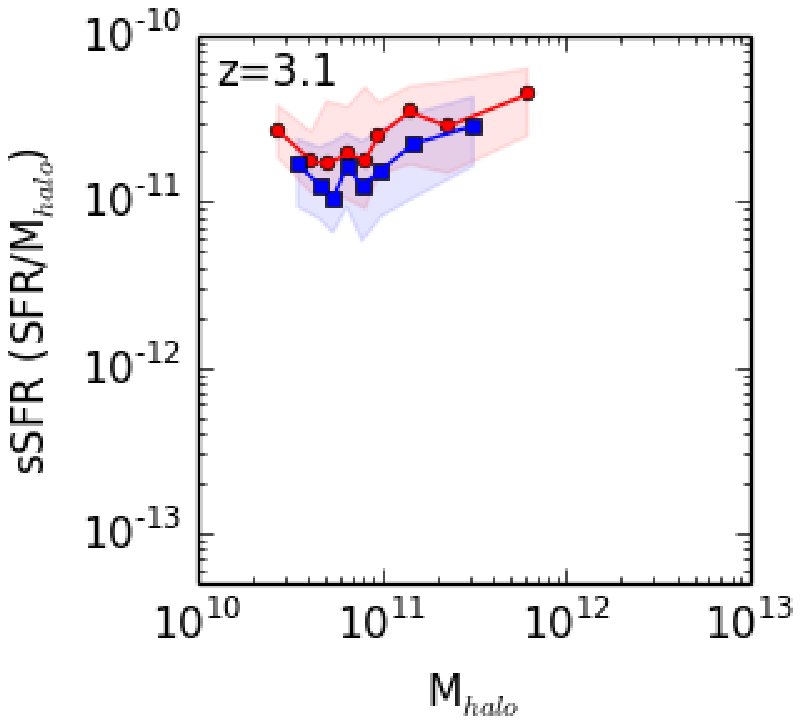}
\includegraphics[scale=0.79,trim=9mm 0mm 0mm 0mm, clip]{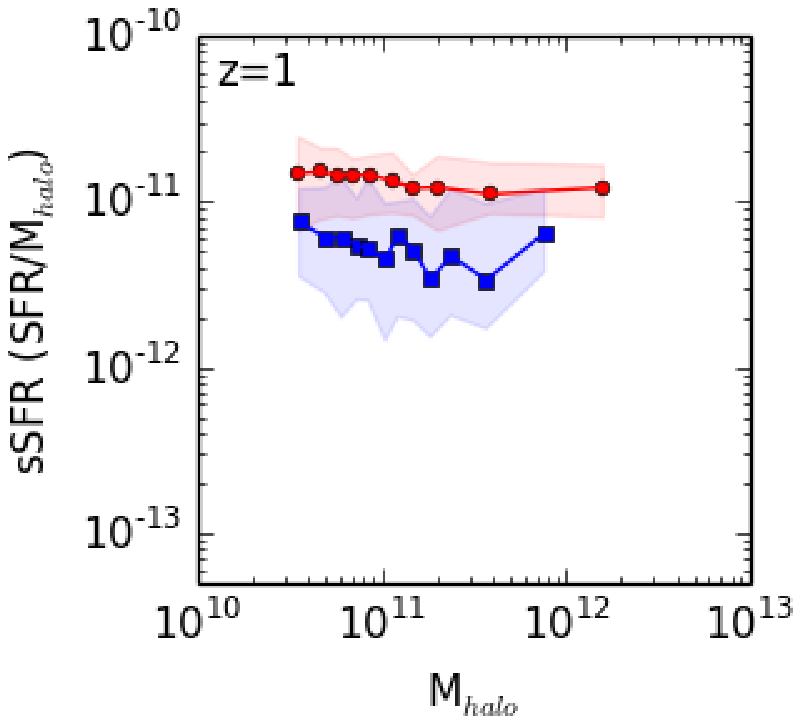}
\includegraphics[scale=0.79,trim=9mm 0mm 0mm 0mm, clip]{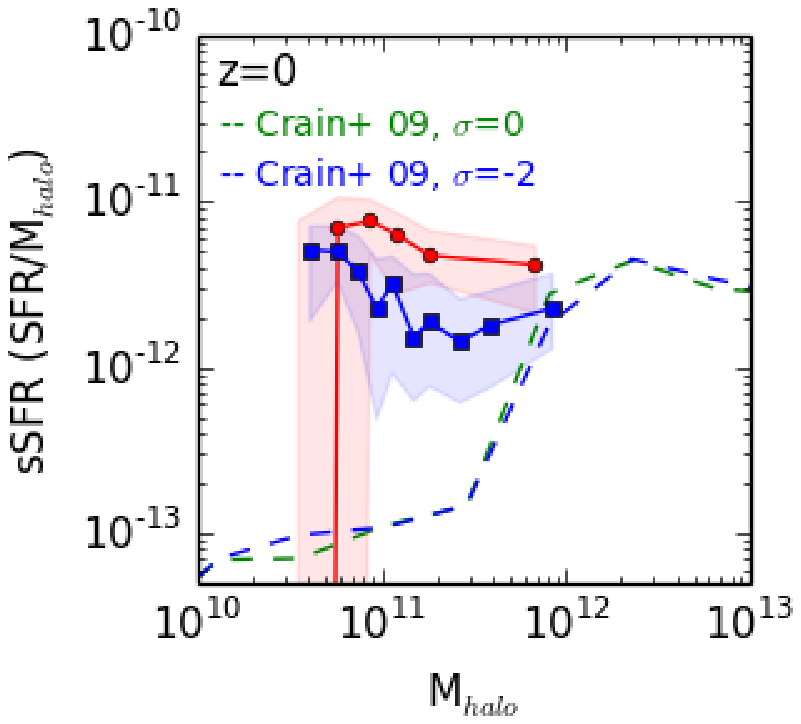}
%\includegraphics[scale=0.79,trim=9mm 0mm 0mm 0mm, clip]{../Mdm_med_plots_notsat2gals_nomean_sSFRdmonly_0000_carefulssfr_notnearedge_forpaper.png}
%\includegraphics[scale=0.72,trim= 8mm 6mm 80mm 78mm, clip]{Mdm_med_plots_notsat2gals_nomean_sSFRdm_0000_carefulssfr_notnearedge.png}
%\end{center}
\caption{The halo sSFR (SFR/M$_{DM halo}$) as a function of dark matter halo mass.  Each point is the median value of 50 galaxies binned by halo mass.  The shaded region shows the 25-75\% range of values.  The red denotes the C box galaxies, and the blue denotes the V box galaxies.  We overplot results from Crain et al. 2009 in the $z$=0 panel for comparison.  From left to right the redshifts are $z$=3.1, 1.0, 0.  The halo sSFR tends to be higher for galaxies in the overdensity (C box) than in the underdensity (V box).  As the redshift decreases, the difference between C box galaxies and V box galaxies become more dramatic.  See discussion in Section \ref{sec:comparison}.}\label{fig-hsSFR}
\end{figure*}

Much of the work performed examining how environment, often measured through clustering, affects galaxy properties has been performed using SAMs.  A dark matter-only N-body simulation is used and galaxies are assumed to populate halos from early times.  The assembly history of the halos, such as smooth accretion and mergers, is also assigned to the galaxy.  For example, the accretion of gas follows the accretion of dark matter using the global baryonic fraction, then cools at a prescribed rate until it can form stars.  The internal processes of galaxies, such as gas cooling, star formation and feedback processes, are generally simplified prescriptions that are tuned so that the $z$=0 population matches observed galaxy populations (e.g. as described in Jung, Lee \& Yi 2014).  

Using such a SAM, Jung et al. (2014) find a small difference in regions of different large-scale (7 h$^{-1}$ Mpc) density, specifically that low-mass ($<$10$^{12}$ M$_{\odot}$) halos have slightly higher stellar masses in high-density environments. 

Our results qualitatively agree with those of Jung et al. (2014), although they find a much smaller difference in the stellar mass of galaxies in overdense versus underdense regions, and only for low-mass halos (10$^{11}$-10$^{11.5}$ M$_{\odot}$/$h$).  They show that in their simulation this is because there is more cold gas in low-mass halos in higher density environments.  We have also posed this as an explanation for our different stellar masses.  However, we see a much larger effect for a much broader range of galaxy masses.  We consider a few possible explanations.  First, Jung et al. (2014) considers environment on the scale of 7 $h^{-1}$ Mpc while we are considering environment measured on $\sim$25 Mpc scale.  Although Jung et al. (2014) tested that scales from 3-9 $h^{-1}$ Mpc did not effect their results, Gao et al. (2005) found that assembly bias increases with increasing length scale, and Jung et al. (2014) do not extend to the scale we use to define environment.  Secondly, we directly calculate the cooling rate using the gas chemistry, and perhaps gas is more metal-rich in the overdense environment and so can cool more quickly.  Indeed, Cen (2013) finds that gas flowing into halos (with negative radial velocities) tends to have higher metallicity in the overdense C box than in the underdense V box.  It is possible that metals can escape lower density filaments more easily therefore resulting in less metal-enrichment of galaxies in low density environments.  This is an extension of the discussion of metal enrichment of clusters via filaments by Rasmussen \& Ponman (2009).  However, we suspect that the most important difference is that we directly calculate any hydrodynamical effects from neighbors.  As we discuss in Section \ref{sec:story}, we believe that one driver of our results is interactions with nearby galaxies (most of which are not satellites) driving gas instabilities and inflows that increase SFRs.  This is a process that is not treated in SAMs.

We can also compare with the other hydrodynamical simulation that compares galaxies in different environments, and find that our results differ from those in Crain et al. (2009).  The GIMIC simulations use GADGET3 to hydrodynamically ``re"-simulate regions of the Millennium simulation at (-2, -1, 0, +1, +2)$\sigma$ (on the scale of $\sim$20 Mpc) of the mean density to determine what drives differences in the SFRD.  While the peaks of their curves are between $z$=2-3, in agreement with our results, they find a smaller difference in the normalization of the SFRD curves in the high- and low-density environments.  

Other properties of their galaxy populations differ from ours.  For example, Crain et al. (2009) do not find much difference in either the shape or the magnitude of the stellar mass functions in their over dense versus under dense regions.  However, while they reproduce the overall number of galaxies more massive than 10$^9$ M$_{\odot}$, their stellar mass functions do not match the observed shape of the $z$=0 stellar mass function of Li \& White (2009) (as discussed in their Figure 3).  We note that this group has since run the `Evolution and Assembly of GaLaxies and their Environments' (EAGLE) simulation project, which has an excellent fit to the $z$=0 galaxy stellar mass function (Schaye et al. 2015).  However, the EAGLE simulations do not compare different large scale environments, so we cannot compare our work to these simulations.  In Figure 3 of Cen (2011) the luminosity function of galaxies in the combined C and V boxes is shown to be a good fit to Blanton et al. (2003), although the highest luminosity, i.e. most massive, galaxies require a post-processing addition of AGN feedback to match observations.  In addition, Kreckel et al. (2011) focus only the V box, and reproduce the luminosity function of void galaxies found by Hoyle et al. (2005).  The void luminosity function differs from the total luminosity function both in that the normalization of the low luminosity end drops by more than an order of magnitude and the knee of the function shifts.  

In this paper we track galaxies over time rather than look at the galaxy population at individual redshifts.  However, in order to determine whether we would get the same results, in Figure \ref{fig-hsSFR} we make a halo sSFR figure similar to Figure 8 in Crain et al. (2009).  Here we look at $z$=3.1,1.0, and 0.  We bin our galaxies by halo mass, with 50 galaxies in a bin, except the highest mass bin has between 50-100 galaxies.  In red we plot the C box galaxies and in blue we plot the V box galaxies.  The solid lines and symbols are the median values and the shaded region denotes the central 25-75\% of values.  We briefly note that the low mass galaxies at $z$=0 have masses below our lowest mass bin in Figures \ref{fig-hmbhm}-\ref{fig-neighbors}, and most of these galaxies are not tracked even to $z$=1.  The precipitous drop in the sSFR is quite probably due to poor resolution of these galaxies, or may possibly indicate that the environment is dramatically affecting the SFRs of these galaxies beyond two virial radii of their nearest neighbor or that these are splashback galaxies.  The dashed lines at $z$=0 show some of the Crain et al. (2009) data.  First, our halo sSFR agrees well with the high mass end of the halo sSFR in Crain et al. (2009), but is relatively flat with halo mass and does not drop at lower masses.  We speculate that this is because we do not include kinetic feedback as in the GIMIC simulations, so we do not have a sudden drop below about 10$^{12}$ M$_{\odot}$ from gas blow-out.  Unlike Crain et al. (2009), we find that at lower redshifts the halo sSFR is clearly different in differing environments, across all of the halo masses we probe.  Even at $z$=3.1, we see a hint that the halo sSFRs of galaxies in the overdense environment is larger than that of galaxies in the underdense environment.  At earlier redshifts, the sSFRS in the two environments are indistinguishable. 

The reason for these differences is difficult to determine as the simulations are quite different.  Crain et al. (2009) use GADGET3, a smooth-particle hydrodynamics code.  Their gravitational softening length is fixed in physical space at $z$$\le$3 to 1 $h^{-1}$ kpc in the intermediate resolution runs (from which most of their results are drawn), which is twice our resolution scale.  Their cosmology is matched to the Millennium Simulation, so is also different from ours.  As discussed above, they also employ kinetic rather than thermal feedback.    

\section{Conclusions}\label{sec:conclusions}

In this paper we have used a cosmological hydrodynamical simulation to examine the stellar mass to halo mass ratio of central galaxies in two different large-scale ($\sim$20 Mpc) environments.  When we focus on the growth of dark matter halos (Figure \ref{fig-hmbhm}), we reproduce the assembly bias found by Gao et al. (2005)--galaxy halos in overdense environments have earlier formation times than those in underdense environments.  

Importantly, we have found that the halo mass alone cannot determine the SMHM ratio of galaxies.  Specifically, in the large-scale overdensity central galaxies with halo masses from 10$^{11}$-10$^{12.9}$ M$_{\odot}$ have higher SMHM ratios than galaxies in a large-scale underdensity (Figure \ref{fig-hmbsmhmr}).  Even when we force the number of galaxies within 2 pMpc at $z$=0 in the overdense regions to be similar to or smaller than the number of surrounding galaxies in the underdense region, we cannot erase this difference.  

We posit that this result is due to the earlier formation times of halos in a large-scale overdensity and to the larger number of nearby galaxies and filaments at early times.  Because the halos in overdense environments are more massive at early times, they can accrete gas and therefore form stars more quickly than those in underdense environments.  Although at later times the V box halos grow more quickly than the C box halos, by that time there is generally less gas accretion, so SFRs are necessarily lower and the SMHM ratio does not equalize.  In addition, particularly at early times, halos in overdense environments have a larger number of neighbors, which can drive instabilities in the gas and increase the SFR of central galaxies.  Finally, there are more filaments feeding galaxies in higher density environments, particularly at early times.

We find that the $z$=0 sSFR of galaxies is higher in the underdense region, in agreement with observational results (Rojas et al. 2004; Grogin \& Geller 1999,2000; Szomoru et al. 1996).  To test our results more robustly, we need more detailed observations of the SFR and stellar masses of a large sample of galaxies in deep, large voids.  In particular, more comparisons of the Tully-Fisher relation using red bands to determine stellar mass would be useful.  One difficulty will be to make sure we are comparing only central galaxies in both underdense and overdense environments. 

Different SMHM ratios in different large-scale environments means that assigning stellar masses based entirely on halo masses in simulations may not produce a realistic galaxy population, and including the $z$=0 local density in the stellar mass assignment will not compensate for the different environments in which the galaxies previously evolved.

\acknowledgements
The authors would like to thank Dr. Claire Lackner for the use of her merger trees, and Dr. Robert Crain for sharing his data.  They would also like to thank the referee for comments that greatly improved the clarity of the paper.  Computing resources were in part provided by the NASA High-
End Computing (HEC) Program through the NASA Advanced
Supercomputing (NAS) Division at Ames Research Center.
The research is supported in part by NSF grant AST-1108700 and NASA grant NNX12AF91G.  ST was supported by the Lyman Spitzer, Jr. Postdoctoral Fellowship and the Alvin E. Nashman Fellowship in Theoretical
Astrophysics.

\end{document}